\documentclass[twocolumn,twoside,aps,superscriptaddress]{itpprepr}

\usepackage{amsfonts}
\usepackage{graphicx}
\usepackage{hyperref}
\usepackage{url}

\newcommand{\ave}[1]{\left\langle{#1}\right\rangle}
\newcommand{\bigo}[1]{\mathcal{O}(#1)}
\newcommand{\binomial}[2]{{{#1} \choose {#2}}}

\newcommand{\Tr}{\mathop{\rm Tr}}
\newcommand{\mod}{\mathop{\rm mod}\nolimits}

\newcommand{\mat}[1]{\ensuremath{\mathbf{#1}}}
\newcommand{\potts}[1]{\vec{e}^{\,(#1)}}

\newcommand{\mdaddr}{Inst.\ f.\ Theor.\ Physik, Univ.\ Magdeburg, Universit\"atsplatz 2, 39106 Magdeburg, Germany}

\begin{document}

\title{Phase Transition in Multiprocessor Scheduling}

\author{Heiko Bauke} \email[E-mail:]{heiko.bauke@student.uni-magdeburg.de}
\affiliation{\mdaddr} \author{Stephan Mertens}
\email[E-mail:]{stephan.mertens@physik.uni-magdeburg.de} \affiliation{\mdaddr}
\affiliation{The Abdus Salam International Centre of Theoretical Physics,
  St.~Costiera 11, 34100 Trieste, Italy} \author{Andreas Engel}
\email[E-mail:]{andreas.engel@physik.uni-magdeburg.de} \affiliation{\mdaddr}

\date{\today}

\begin{abstract}
  The problem of distributing the workload on a parallel computer to minimize
  the overall runtime is known as \textsc{Multiprocessor Scheduling Problem}.
  It is NP-hard, but like many other NP-hard problems, the average hardness of
  random instances displays an ``easy-hard'' phase transition. The transition
  in \textsc{Multiprocessor Scheduling} can be analyzed using 
  elementary notions from
  crystallography (Bravais lattices) and statistical mechanics (Potts
  vectors). The analysis reveals the control parameter of the transition and
  its critical value including finite size corrections.  The transition is
  identified in the performance of practical scheduling algorithms.
\end{abstract}


\maketitle

One of the major problems in parallel computing is load balancing, the even
distribution of the workload on the processors of a parallel computer. The
goal is to find a \emph{schedule}, i.e.\ an assignment of $N$ tasks to $q$
processors so as to minimize the largest task finishing time (makespan). This
is a very hard problem, since individual tasks may depend on each other and
the time a task runs on a given processor may not be known in advance.

The most simple variant of a scheduling problem is known as
\textsc{Multiprocessor Scheduling Problem, Msp} \cite{ausiello:etal:99}. Here
the $N$ tasks are independent, all $q$ processors are equal and the running
time $a_i\in\mathbb{N}$ of each task is known in advance.  A schedule is a map
$s: \{1,\ldots,N\}\mapsto\{1,\ldots,q\}$ with $s_i=\alpha$ denoting that task
$i$ is assigned to processor $\alpha$. The problem then is to minimize the
makespan
\begin{equation}
  \label{eq:def-T}
  T(s_1,s_2,\ldots,s_N) = \max_{\alpha}\left\{A_\alpha = 
    \sum_{i=1}^N a_{i} \delta(s_i-\alpha)\right\}\,.
\end{equation}
$A_\alpha$ is the total workload of processor $\alpha$ and $\delta$ is the
usual Kronecker symbol. 
\textsc{Msp} belongs to the class of NP-hard optimization 
problems \cite{garey:johnson:79,mertens:02a},
which
basically means that no one has ever found a scheduling algorithm that is
significantly faster than exhaustive search through all $\bigo{q^N}$
schedules, and probably no one ever will. With computational costs that
increase exponentially with the problem size $N$, \textsc{Msp} must be
considered intractable.

NP-hardness refers to worst case scenarios. On the other hand, \emph{randomly
  generated instances} of NP-hard problems often are surprisingly easy to
solve, and \textsc{Msp} is no exception. Numerical experiments
\cite{gent:walsh:96} with $a_i$ being random $B$-bit integers reveal two
distinct regimes: For small values of $\kappa=B/N$ typical instances can be
solved without exponential search, for large values of $\kappa$, exponential
search is mandatory. The transition from the ``easy'' to the ``hard''
transition gets sharper as $N$ increases, and in the limit $N\to\infty$ it
happens at a threshold value $\kappa_c(q)$. Similar phase transitions have
been observed in many other combinatorial optimization problems
\cite{cheeseman:etal:91}. A transition in the typical algorithmic complexity
normally corresponds to a structural change in the typical instances of the
random ensemble. The latter in turn can be analyzed using the methods and
notions from statistical mechanics \cite{tcs-phasetransitions}. An outstanding
example for the fruitfulness of this interdisciplinary approach is the
ana\-lysis of random $K$-\textsc{Satisfiability} with the replica
\cite{monasson:etal:99} and the cavity method
\cite{zecchina:parisi:mezard:02}.

The structural change in \textsc{Msp} that lies underneath the algorithmic
transition is the appearance of {\em perfect} schedules.  Let $r = \sum_j a_j
\mod q$. A schedule with
\begin{equation}
  \label{eq:perfect-schedule}
  A_\alpha - \left\lfloor\frac{1}{q}\sum_{i=1}^Na_j\right\rfloor = \left\{
    \begin{array}{ccl}
      1 & \text{if} & 1 \leq \alpha \leq r \\
      0 & \text{if} & r < \alpha \leq q
    \end{array}
  \right.
\end{equation}
(and its $\binomial{q}{r}$ equivalent rearrangements) is called perfect since
it obviously minimizes the makespan $T$ (Eq.~\ref{eq:def-T}). Whenever an
algorithm runs into a perfect schedule it can stop the search possibly before
having explored an exponential part of the search space. Numerical simulations
in fact indicate that the probability that a random instance has a perfect
schedule decreases from $1$ for $\kappa=0$ to $0$ as $\kappa \gg 1$, and for
large $N$ the probability jumps abruptly from $1$ to $0$ as $\kappa$ crosses a
critical value $\kappa_c(q)$. In this letter we will calculate $\kappa_c(q)$.

The \textsc{Msp} for $q=2$ is known as \textsc{Number Partioning Problem, Npp}
\cite{hayes:npp}. The transition point of the \textsc{Npp}, $\kappa_c(2)$, has
been calculated within the canonical formalism of statistical mechanics
\cite{mertens:98a} and the results have been confirmed recently by rigorous
proofs \cite{borgs:chayes:pittel:01}. In principle one could extend the method
of \cite{mertens:98a} to general $q$, but here we will follow a microcanonical
approach.

A first estimate for the critical $\kappa$ can be obtained following a nice
heuristic argument \cite{gent:walsh:96}: Given the values of the $a_i$ are
$\kappa N$-bit integers, the workloads defined by
Eq.~\ref{eq:perfect-schedule} are (neglegting for the moment carry bits) also
$\kappa N$-bit integers. The probability that a randomly chosen schedule
$(s_1,s_2,\ldots,s_N)$ realizes a particular value of $A_1$ is therefore
$2^{-\kappa N}$. Neglecting correlations the chance to realize all the
workloads defined in Eq.~\ref{eq:perfect-schedule} is hence $2^{-(q-1)\kappa
  N}$ ($A_q$ is fixed implicitly). Since there are $q^N$ different schedules
the number of perfect ones is roughly given by $q^N 2^{-(q-1)\kappa N}$.  The
occurrence of a phase transition in \textsc{Msp} is now easily understood.  If
$\kappa$ is small we expect an exponential number of perfect solutions, but
for $\kappa$ larger than a critical value
\begin{equation}
  \label{eq:kappa_c-annealed}
  \kappa_c = \frac{\log_2 q}{q-1}
\end{equation}
the number of perfect solutions is exponentially small.

\begin{figure}[b]
  \centering \includegraphics[width=0.8\columnwidth]{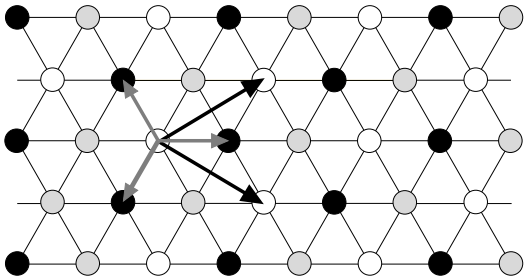}
  \caption{Lattice of target vectors for $q=3$ with the three Potts
    vectors (gray) and the two primitive vectors $\vec{b}_\alpha$ (black).
    The white (gray, black) lattice points correspond to $\sum a_j \mod 3 = 0
    (1,2)$.}
  \label{fig:bravais}
\end{figure}

The first step of our approach is a convenient encoding of the schedules and
the cost function. The workloads on the processors are not independent:
what is removed from one processor must be done by another. We can incorporate
this constraint automatically by encoding the schedule as Potts vectors
\cite{wu:82}.  Potts vectors $\potts{\alpha}$ are $(q-1)$-dimensional unit
vectors pointing at the $q$ corners of a $(q-1)$-dimensional hypertetrahedron
(see Fig.~\ref{fig:bravais} for the case $q=3$). This implies that the angle
between two Potts vectors is the same for all pairs of different vectors,
\begin{equation}
  \label{eq:inner-product}
  \potts{\alpha} \cdot \potts{\beta} = \frac{q\delta(\alpha-\beta)-1}{q-1}.
\end{equation}
A schedule is encoded by $N$ Potts vectors $\vec{s}_j$, where
$\vec{s}_j=\potts{\alpha}$ means that task $j$ is assigned to processor
$\alpha$, i.e.
\begin{equation}
  \label{eq:def-s_j}
  \vec{s}_j=\sum_\beta \delta(s(j)-\beta) \potts{\beta}.
\end{equation}
The {\em target vector}
\begin{equation}
  \label{eq:def-target}
  \vec{E}(\{\vec{s}\}) = \sum_{j=1}^N a_j \vec{s}_j
\end{equation}
encodes the workload of all processors,
\begin{equation}
  A_\alpha - \frac{1}{q}\sum_ja_j = \frac{q-1}{q} \vec{E}\cdot\potts{\alpha},
\end{equation}
and minimizing $T$ (Eq.~\ref{eq:def-T}) is equivalent to minimizing $\vec{E}$
with respect to the supremum norm \cite{mattis}.

For integer values $a_i$ the minimal change of a schedule is to remove 1 from
one processor and add it to one of the other $q-1$ processors. Hence possible
values of $\vec{E}(\{\vec{s}\})$ are points on a $(q-1)$-dimensional
Bravais-lattice with primitive vectors
\begin{equation}
  \label{eq:def-b}
  \vec{b}_\alpha = \potts{\alpha} - \potts{q} \qquad \alpha=1,\ldots,q-1\,.
\end{equation}
These primitive vectors span a sublattice of the lattice generated by $q-1$
Potts vectors. The sublattice contains every $q$th point of the Potts lattice
and correspondingly there are $q$ classes of such lattice points depending on
$\sum a_j\mod q$ (Fig.~\ref{fig:bravais}).  The volume $V(q)$ of the primitive
cell in our sublattice can be calculated from the Gram determinant,
\begin{equation}
  \label{eq:V}
  V^2(q) = \det(\vec{b}_\alpha \cdot \vec{b}_\beta) = 
  \frac{q^q}{(q-1)^{q-1}}\,.
\end{equation}


\noindent
The average number $\Omega$ of schedules with target $\vec{E}$ is
\begin{equation}
  \label{eq:Omega-1}
  \Omega({\vec{E}}) = \Tr_{\{\vec{s}\}} \ave{\delta\left(\vec{E}- 
                        \sum_{j=1}^N a_j\vec{s}_j\right)} 
\end{equation}
where $\ave{\cdot}$ denotes averaging over the i.i.d.\ random numbers $a_i$.
For fixed schedule $\{\vec{s}_j\}$ and large $N$, the sum $\sum_{j=1}^N
a_j\vec{s}_j$ is Gaussian with mean
 \begin{equation}
  \label{eq:E-mean}
  \Big\langle\vec{E}\Big\rangle = \ave{a}\sum_j \vec{s}_j =: \ave{a} \vec{M}
\end{equation}
and variance of the components $\vec{E}=(E_1,\ldots,E_{q-1})$
\begin{equation}
  \label{eq:E-var}
  \ave{E_i E_k} - \ave{E_i} \ave{E_k} = 
  \sigma_a^2 \sum_j (\vec{s}_j)_i (\vec{s}_j)_k =: \sigma_a^2 g_{i,k}
\end{equation}
with $\sigma_a^2 = \langle a^2\rangle - \langle a \rangle^2$.
``Magnetization'' $\vec{M}$ and variance matrix $\mat{g} = (g_{i,k})$ depend
on the schedule only through the numbers $N_\alpha = \sum_j
\delta(s_j-\alpha)$ of tasks assigned to processor $\alpha$:
\begin{equation}
  \label{eq:E-mean-2}
  \vec{M} = \sum_{\alpha=1}^q N_\alpha \potts{\alpha} \qquad
  g_{i,k} = \sum_{\alpha=1}^q N_{\alpha} e^{(\alpha)}_i e^{(\alpha)}_k.
\end{equation}
The trace over $\{\vec{s}\}$ is basically an average over all trajectories of
a random walk in $q-1$ dimensions. For large $N$ this average is dominated by
trajectories with $\vec{M} = 0$, i.e.\ $N_\alpha = N/q$. For these
trajectories, the matrix \mat{g} is diagonal,
\begin{equation}
  \label{eq:diag-g}
  g_{i,k} = \frac{\delta(i-k)}{q-1},
\end{equation}
and we have basically an independent random walk in each of the $q-1$
directions of our lattice.  The probability to occupy after $N$ steps a
position $E_\alpha$ away from the origin in direction $\alpha$ is therefore
given by
\begin{equation}
  \label{eq:p-Ea}
  p(E_\alpha) = \frac{\sqrt{q-1}}{\sqrt{2\pi N\ave{a^2}}} \exp\left(-\frac{(q-1) E_\alpha^2}{2 N\ave{a^2}}\right).
\end{equation}
The probability of finding the $q-1$ walkers at $\vec{E}$ reads
\begin{equation}
  \label{eq:p-E}
  p(\vec{E}) = \left(\frac{(q-1)}{2\pi N\ave{a^2}}\right)^{(q-1)/2} \exp\left(-\frac{(q-1)|\vec{E}|^2}{2 N\ave{a^2}}\right).
\end{equation}    
To get the number of schedules with given target vector $\vec{E}$ we have to
multiply the density $p(\vec{E})$ by $q^N$ and the volume $V(q)$ of the
primitive cell of our Bravais lattice,
\begin{equation}
  \label{eq:omega}
   \Omega(\vec{E}) = \frac{q^N q^{q/2}}{\Big(2\pi N\ave{a^2}\Big)^{(q-1)/2}} \exp\left(- \frac{q-1}{2N\ave{a^2}} |\vec{E}|^2 \right).
\end{equation}
For target vectors $|\vec{E}| = \bigo{1}$ the distribution looks
essentially flat as $N\to\infty$, i.e.\ there are as many perfect
schedules as there are suboptimal schedules.

The density of \emph{scalar} quantities like $|\vec{E}|^2$ gets a factor
$|\vec{E}|^{q-2}$ from the volume element in $(q-1)$-dimensional spherical
coordinates. For $q > 2$ this leads to a maximum of the microcanonical entropy
at some value $|\vec{E}|>0$, and this maximum gets sharper with increasing
$q$, a scenario that has been observed in Monte Carlo simulations
\cite{lima:menezes:01}.  However, this implies no fundamental difference
between $q=2$ and $q>2$ as claimed in \cite{lima:menezes:01} but is of purely
geometrical origin.

For the location of the phase transition we can concentrate on perfect
schedules, i.e.\ we set $|\vec{E}| = \bigo{1}$ and we assume that the $a_i$'s
are uniformely distributed $\kappa N$-bit integers.  From
\begin{equation}
  \label{eq:insert-kappa}
  \ave{a^2} = 
  \frac{1}{3} \, 2^{2\kappa N} \left(1 - \bigo{2^{-\kappa N}}\right)
\end{equation}
we get
\begin{equation}
  \label{eq:entropy}
  \log_2\Omega(0) = N (q-1)\cdot (\kappa_c - \kappa)
\end{equation}
with
\begin{equation}
  \label{eq:kappa_c}
  \kappa_c = \frac{\log_2 q}{q-1} - \frac{1}{2N}
  \log_2\left(\frac{2\pi N}{3 q^{q/(q-1)}}\right).
\end{equation}
The first term corresponds to the result of the heuristic argument from above
(Eq.~\ref{eq:kappa_c-annealed}). The second term represents finite size
corrections.  For $q=2$ Eqs.~\ref{eq:entropy} and \ref{eq:kappa_c} reduce to
the known results for number partitioning
\cite{mertens:98a,borgs:chayes:pittel:01}.

\begin{figure}
  \centering \includegraphics[width=0.95\columnwidth]{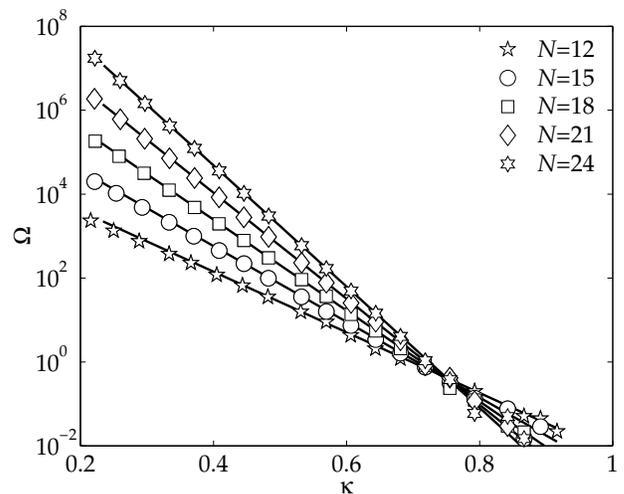}
  \caption{Numerical measurements of $\Omega$ for $q=3$. Symbols are averages
    over $10^3$ random instances with $\sum_j a_j \mod q = 0$, lines are given
    by Eq.~\ref{eq:entropy}. The errorbars of the enumeration data are smaller
    than the symbols. }
  \label{fig:omega}
\end{figure}

\begin{figure}
  \centering \includegraphics[width=0.95\columnwidth]{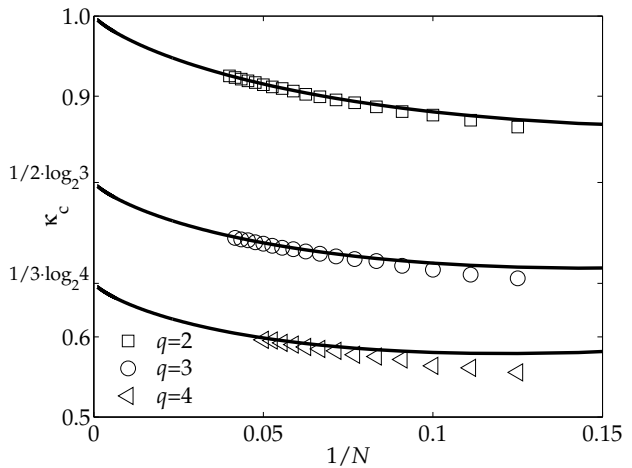}
  \caption{$\kappa_c$ from linear regression of the numerical data
    for $\Omega$ (symbols) compared to Eq.~\ref{eq:kappa_c} (curves)}
  \label{fig:kappa_c}
\end{figure}

According to Eq.~\ref{eq:entropy} the microcanonical entropy
$\log\Omega(\vec{E})$ is a linear function of $\kappa$ for large $N$.  In fact
this linearity already holds for rather small values of $N$, as can be seen
from numerical enumerations (Fig.~\ref{fig:omega}). Linear regression on the
data for $\log\Omega(\vec{E})$ gives numerical values for $\kappa_c(N)$.
These values in turn agree well with the predictions of Eq.~\ref{eq:kappa_c}
for larger values of $N$ (Fig.~\ref{fig:kappa_c}).

Up to this point we have discussed \emph{static} properties of random
\textsc{Msp}. How do they affect the \emph{dynamical} behavior of search
algorithms? An obvious algorithm is to sort the tasks $a_i$ in decreasing
order and to assign the first (and largest) task to processor $1$. The next
tasks are each assigned to the processor with the smallest total workload so
far. Proceed until all tasks are assigned.  Ties are broken by selecting the
processor with the lower rank.  This so called greedy heuristics usually
produces poor schedules, but it can be extended to an algorithm that yields
the optimum schedule. Instead of assigning a task to one processor, the
extended algorithm branches: in the first branch it follows the heuristic rule
and assigns the task to the processor with the lowest workload, in the second
branch it selects the processor with the second lowest workload and so on.
Ignoring ties this algorithm generates the complete search tree with its $q^N$
leaves and hence will eventually find the true optimum. This algorithm is
known as Complete Greedy Algorithm (CGA) \cite{korf:98}.  Of course it is
worst-case exponential, but with pruning we can hope to achieve a speedup for
the typical case.  The most efficient pruning rule is to simply stop the
moment one hits upon a perfect schedule.  A less efficient pruning rule
applies if the sum of the unassigned tasks is smaller than the difference
between the current maximal and minimal workloads.  In this case one can
simply assign all remaining tasks to the processor with the minimum workload.

\begin{figure}[htbp]
  \centering \includegraphics[width=0.95\columnwidth]{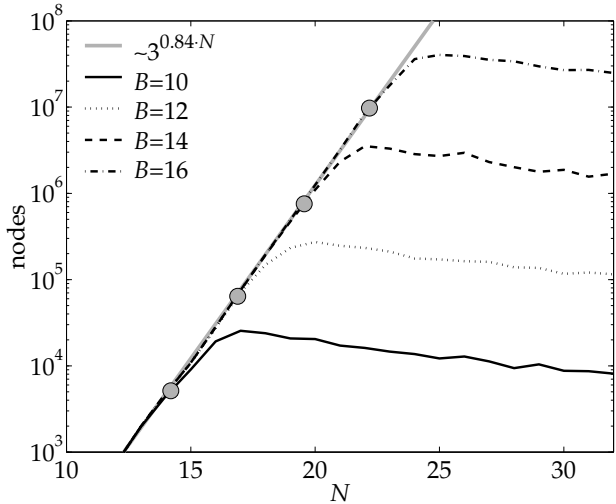}
  \caption{Median of number of nodes traversed by the
    complete greedy algorithm for $q=3$ and fixed number of bits $B$.  The
    circles mark the critical system size $N_c$ given by Eq.~\ref{eq:Nc}}
  \label{fig:nodes}
\end{figure}

In our simulations we fix the number $B$ of bits in the $a_i$'s and measure
the number of nodes traversed by CGA as a function of $N$.  In this case
$\kappa_c$ translates into a critical value $N_c$ for the system size, $N_c$
being the solution of
\begin{equation}
  \label{eq:Nc}
  \frac{B}{N_c} = \frac{\log_2 q}{q-1} - \frac{1}{2N_c}
  \log_2\left(\frac{2\pi N_c}{3 q^{q/(q-1)}}\right).
\end{equation}
Fig.~\ref{fig:nodes} shows a typical result for $q=3$.  For $N < N_c$ the
number of nodes traversed by CGA increases like $3^{xN}$ with $x\approx0.84$.
The fact that $x < 1$ is due to the pruning. As soon as $N > N_c$, the pruning
by perfect solutions takes effect and the growth slows down significantly.
Eventually the search costs even decrease with increasing $N$. This indicates
that the algorithm takes advantage of the growing number of perfect solutions,
allthough their relative number still decays exponentially.  There are
algorithms that outperform simple CGA, but the differences show only for $N >
N_c$: With better algorithms the subexponential growth and the decrease of the
search costs set in at values of $N$ closer to but always above $N_c$
\cite{korf:98}.

To conclude we have shown that \textsc{Multiprocessor Scheduling} has a phase
transition controlled by the numerical resolution
(number of bits) of the
individual task sizes. The
``easy'' phase is characterized by an exponential number of perfect schedules,
the ``hard'' phase by the absence of perfect schedules.  Note that for $q=2$
it has been demonstrated that deep in the ``hard'' phase the system behaves
essentially like a random energy model \cite{mertens:00a}. This fact allowed
the calculation of the complete statistics of the optimal solutions and it
explains the bad performance of heuristic algorithms. It would be interesting
to extend this approach to $q > 2$.

\acknowledgments Discussions with Ido Kanter are gratefully acknowledged.
S.M.~enjoyed the hospitality of the ICTP, Trieste.  All numerical simulations
have been done on our Beowulf cluster \textsc{Tina}\cite{tina:url}.

\bibliographystyle{apsrev} 
\bibliography{complexity,cs,q-npp}

\begin{thebibliography}{19}
\expandafter\ifx\csname natexlab\endcsname\relax\def\natexlab#1{#1}\fi
\expandafter\ifx\csname bibnamefont\endcsname\relax
  \def\bibnamefont#1{#1}\fi
\expandafter\ifx\csname bibfnamefont\endcsname\relax
  \def\bibfnamefont#1{#1}\fi
\expandafter\ifx\csname citenamefont\endcsname\relax
  \def\citenamefont#1{#1}\fi
\expandafter\ifx\csname url\endcsname\relax
  \def\url#1{\texttt{#1}}\fi
\expandafter\ifx\csname urlprefix\endcsname\relax\def\urlprefix{URL }\fi
\providecommand{\bibinfo}[2]{#2}
\providecommand{\eprint}[2][]{\url{#2}}

\bibitem[{\citenamefont{Ausiello et~al.}(1999)\citenamefont{Ausiello,
  Crescenzi, Gambosi, Kann, Marchetti-Spaccamela, and
  Protassi}}]{ausiello:etal:99}
\bibinfo{author}{\bibfnamefont{G.}~\bibnamefont{Ausiello}},
  \bibinfo{author}{\bibfnamefont{P.}~\bibnamefont{Crescenzi}},
  \bibinfo{author}{\bibfnamefont{G.}~\bibnamefont{Gambosi}},
  \bibinfo{author}{\bibfnamefont{V.}~\bibnamefont{Kann}},
  \bibinfo{author}{\bibfnamefont{A.}~\bibnamefont{Marchetti-Spaccamela}},
  \bibnamefont{and} \bibinfo{author}{\bibfnamefont{M.}~\bibnamefont{Protassi}},
  \emph{\bibinfo{title}{Complexity and Approximation}}
  (\bibinfo{publisher}{Spring\-er-Verlag}, \bibinfo{address}{Ber\-lin
  Hei\-del\-berg New~York}, \bibinfo{year}{1999}).

\bibitem[{\citenamefont{Garey and Johnson}(1997)}]{garey:johnson:79}
\bibinfo{author}{\bibfnamefont{M.~R.} \bibnamefont{Garey}} \bibnamefont{and}
  \bibinfo{author}{\bibfnamefont{D.~S.} \bibnamefont{Johnson}},
  \emph{\bibinfo{title}{Computers and Intractability. A Guide to the Theory of
  {NP}-Completeness}} (\bibinfo{publisher}{W.H.~Freeman}, \bibinfo{address}{New
  York}, \bibinfo{year}{1997}).

\bibitem[{\citenamefont{Mertens}(2002)}]{mertens:02a}
\bibinfo{author}{\bibfnamefont{S.}~\bibnamefont{Mertens}},
  \bibinfo{journal}{Computing in Science and Engineering}
  \textbf{\bibinfo{volume}{4}}, \bibinfo{pages}{31} (\bibinfo{year}{2002}).

\bibitem[{\citenamefont{Gent and Walsh}(1996)}]{gent:walsh:96}
\bibinfo{author}{\bibfnamefont{I.~P.} \bibnamefont{Gent}} \bibnamefont{and}
  \bibinfo{author}{\bibfnamefont{T.}~\bibnamefont{Walsh}}, in
  \emph{\bibinfo{booktitle}{Proc.~of ECAI-96}}, edited by
  \bibinfo{editor}{\bibfnamefont{W.}~\bibnamefont{Wahlster}}
  (\bibinfo{publisher}{John Wiley \& Sons}, \bibinfo{address}{New York},
  \bibinfo{year}{1996}), pp. \bibinfo{pages}{170--174}.

\bibitem[{\citenamefont{Cheeseman et~al.}(1991)\citenamefont{Cheeseman,
  Kanefsky, and Taylor}}]{cheeseman:etal:91}
\bibinfo{author}{\bibfnamefont{P.}~\bibnamefont{Cheeseman}},
  \bibinfo{author}{\bibfnamefont{B.}~\bibnamefont{Kanefsky}}, \bibnamefont{and}
  \bibinfo{author}{\bibfnamefont{W.~M.} \bibnamefont{Taylor}}, in
  \emph{\bibinfo{booktitle}{Proc. of IJCAI-91}}, edited by
  \bibinfo{editor}{\bibfnamefont{J.}~\bibnamefont{Mylopoulos}}
  \bibnamefont{and} \bibinfo{editor}{\bibfnamefont{R.}~\bibnamefont{Rediter}}
  (\bibinfo{publisher}{Morgan Kaufmann}, \bibinfo{address}{San Mateo, CA},
  \bibinfo{year}{1991}), pp. \bibinfo{pages}{331--337}.

\bibitem[{\citenamefont{Dubois et~al.}(2001)\citenamefont{Dubois, Monasson,
  Selman, and Zecchina}}]{tcs-phasetransitions}
\bibinfo{editor}{\bibfnamefont{O.}~\bibnamefont{Dubois}},
  \bibinfo{editor}{\bibfnamefont{R.}~\bibnamefont{Monasson}},
  \bibinfo{editor}{\bibfnamefont{B.}~\bibnamefont{Selman}}, \bibnamefont{and}
  \bibinfo{editor}{\bibfnamefont{R.}~\bibnamefont{Zecchina}}, eds.,
  \emph{\bibinfo{title}{Phase Transitions in Combinatorial Problems}}, vol.
  \bibinfo{volume}{265} of \emph{\bibinfo{series}{Theor.\ Comp.\ Sci.}}
  (\bibinfo{year}{2001}).

\bibitem[{\citenamefont{R.Monasson et~al.}(1999)\citenamefont{R.Monasson,
  Zecchina, Kirkpatrick, Selman, and Troyanksy}}]{monasson:etal:99}
\bibinfo{author}{\bibnamefont{R.Monasson}},
  \bibinfo{author}{\bibfnamefont{R.}~\bibnamefont{Zecchina}},
  \bibinfo{author}{\bibfnamefont{S.}~\bibnamefont{Kirkpatrick}},
  \bibinfo{author}{\bibfnamefont{B.}~\bibnamefont{Selman}}, \bibnamefont{and}
  \bibinfo{author}{\bibfnamefont{L.}~\bibnamefont{Troyanksy}},
  \bibinfo{journal}{Nature} \textbf{\bibinfo{volume}{400}},
  \bibinfo{pages}{133} (\bibinfo{year}{1999}).

\bibitem[{\citenamefont{M\'ezard et~al.}(2002)\citenamefont{M\'ezard, Parisi,
  and Zecchina}}]{zecchina:parisi:mezard:02}
\bibinfo{author}{\bibfnamefont{M.}~\bibnamefont{M\'ezard}},
  \bibinfo{author}{\bibfnamefont{G.}~\bibnamefont{Parisi}}, \bibnamefont{and}
  \bibinfo{author}{\bibfnamefont{R.}~\bibnamefont{Zecchina}},
  \bibinfo{journal}{Science}  (\bibinfo{year}{2002}), \bibinfo{note}{published
  online: 10.1126/science.1073287}.

\bibitem[{\citenamefont{Hayes}(2002)}]{hayes:npp}
\bibinfo{author}{\bibfnamefont{B.}~\bibnamefont{Hayes}},
  \bibinfo{journal}{American Scientist} \textbf{\bibinfo{volume}{90}},
  \bibinfo{pages}{113} (\bibinfo{year}{2002}).

\bibitem[{\citenamefont{Mertens}(1998)}]{mertens:98a}
\bibinfo{author}{\bibfnamefont{S.}~\bibnamefont{Mertens}},
  \bibinfo{journal}{Phys.\ Rev.\ Lett.} \textbf{\bibinfo{volume}{81}},
  \bibinfo{pages}{4281} (\bibinfo{year}{1998}).

\bibitem[{\citenamefont{Borgs et~al.}(2001)\citenamefont{Borgs, Chayes, and
  Pittel}}]{borgs:chayes:pittel:01}
\bibinfo{author}{\bibfnamefont{C.}~\bibnamefont{Borgs}},
  \bibinfo{author}{\bibfnamefont{J.}~\bibnamefont{Chayes}}, \bibnamefont{and}
  \bibinfo{author}{\bibfnamefont{B.}~\bibnamefont{Pittel}},
  \bibinfo{journal}{Rand.\ Struct.\ Alg.} \textbf{\bibinfo{volume}{19}},
  \bibinfo{pages}{247} (\bibinfo{year}{2001}).

\bibitem[{\citenamefont{Wu}(1982)}]{wu:82}
\bibinfo{author}{\bibfnamefont{F.}~\bibnamefont{Wu}}, \bibinfo{journal}{Rev.\
  Mod.\ Phys.} \textbf{\bibinfo{volume}{54}}, \bibinfo{pages}{235}
  (\bibinfo{year}{1982}).

\bibitem[{mat()}]{mattis}
\bibinfo{note}{In $L_2$-norm, the minima of $\vec{E}$ correspond to the
  groundstates of a mean-field Potts-anitiferromagnet with Mattis-like
  couplings \cite{mattis:76,vannimenus:mezard:84}.}

\bibitem[{\citenamefont{Lima and de~Menezes}(2001)}]{lima:menezes:01}
\bibinfo{author}{\bibfnamefont{A.}~\bibnamefont{Lima}} \bibnamefont{and}
  \bibinfo{author}{\bibfnamefont{M.}~\bibnamefont{de~Menezes}},
  \bibinfo{journal}{Phys.\ Rev.\ E} \textbf{\bibinfo{volume}{63}},
  \bibinfo{pages}{020106(R)} (\bibinfo{year}{2001}).

\bibitem[{\citenamefont{Korf}(1998)}]{korf:98}
\bibinfo{author}{\bibfnamefont{R.~E.} \bibnamefont{Korf}},
  \bibinfo{journal}{Artificial Intelligence} \textbf{\bibinfo{volume}{106}},
  \bibinfo{pages}{181} (\bibinfo{year}{1998}).

\bibitem[{\citenamefont{Mertens}(2000)}]{mertens:00a}
\bibinfo{author}{\bibfnamefont{S.}~\bibnamefont{Mertens}},
  \bibinfo{journal}{Phys.\ Rev.\ Lett.} \textbf{\bibinfo{volume}{84}},
  \bibinfo{pages}{1347} (\bibinfo{year}{2000}).

\bibitem[{tin()}]{tina:url}
\urlprefix\url{http://tina.nat.uni-magdeburg.de}.

\bibitem[{\citenamefont{Vannimenus and M\'ezard}(1984)}]{vannimenus:mezard:84}
\bibinfo{author}{\bibfnamefont{J.}~\bibnamefont{Vannimenus}} \bibnamefont{and}
  \bibinfo{author}{\bibfnamefont{M.}~\bibnamefont{M\'ezard}},
  \bibinfo{journal}{J.~Physique Lett.} \textbf{\bibinfo{volume}{45}},
  \bibinfo{pages}{L1145} (\bibinfo{year}{1984}).

\bibitem[{\citenamefont{Mattis}(1976)}]{mattis:76}
\bibinfo{author}{\bibfnamefont{D.}~\bibnamefont{Mattis}},
  \bibinfo{journal}{Phys.~Lett.~A} \textbf{\bibinfo{volume}{56}},
  \bibinfo{pages}{421} (\bibinfo{year}{1976}).

\end{thebibliography}

\end{document}